\documentclass[aps,pra,a4paper,reprint,showpacs,nofootinbib,superscriptaddress,floatfix,floats]{revtex4-2}
\usepackage{pifont}
\usepackage{amsmath}
\usepackage{amssymb}
\usepackage{amsfonts}
\usepackage{times,txfonts}
\usepackage{bbold}
\usepackage{bm}
\usepackage{bbm}
\usepackage{dsfont}
\usepackage{hyperref}
\hypersetup{
    colorlinks=true,
    citecolor=blue,
    linkcolor=blue,
    urlcolor=blue
}
\usepackage{color}
\usepackage{graphicx}
\usepackage{float}
\usepackage{epsfig}
\usepackage{verbatim}
\usepackage{epstopdf}
\usepackage{natbib}
\usepackage{ulem}
\usepackage{stackengine} 

\newcommand{\inner}[2]{\left\langle#1\kern-\nulldelimiterspace\left|#2\kern-\nulldelimiterspace\right.\right\rangle}

\DeclareMathOperator{\Tr}{Tr}

\usepackage{afterpage}

\begin{document}

\title{A Real-time Instanton Approach to Quantum Activation}

\author{Chang-Woo Lee}
\thanks{Chang-Woo Lee and Paul Brookes contributed equally to this work.}
\affiliation{School of Computational Sciences, Korea Institute for Advanced Study, 85 Hoegi-ro, Dongdaemun-gu, Seoul 02455, Korea}
\author{Paul Brookes}
\thanks{Chang-Woo Lee and Paul Brookes contributed equally to this work.}
\affiliation{Department of Physics and Astronomy, University College London, Gower Street, London WC1E 6BT, United Kingdom}
\author{Kee-Su Park}
\affiliation{Department of Physics, Sungkyunkwan University, Suwon 16419, Korea}
\author{Marzena H. Szyma\'{n}ska}
\affiliation{Department of Physics and Astronomy, University College London, Gower Street, London WC1E 6BT, United Kingdom}
\author{Eran Ginossar}
\affiliation{Advanced Technology Institute and Department of Physics, University of Surrey, Guildford GU2 7XH, United Kingdom}

\date{\today}

\begin{abstract}
Driven-dissipative nonlinear systems exhibit rich critical behavior, related to bifurcation, bistability and switching, which underlie key phenomena in areas ranging from physics, chemistry and biology \cite{wilhelm_smallest_2009}  to social sciences and economics. The importance of rare fluctuations leading to a dramatic jump between two very distinct states, such as survival and extinction in population dynamics \cite{PhysRevLett.125.048105,Kamenev_2008}, success and bankruptcy in economics \cite{ghashghaie_turbulent_1996} and the occurrence of earthquakes \cite{Knopoff_1977} or of epileptic seizures \cite{Lehnertz_2006}, have been already established. In the quantum domain, switching is of importance in both chemical reactions and the devices used in quantum state detection and amplification \cite{Vijay2009}. In particular, the simplest driven single oscillator model serves as an insightful starting point. Here we describe switching induced by quantum fluctuations and illustrate that an instanton approach within Keldysh field theory can provide a deep insight into such phenomena. We provide a practical recipe to compute the switching rates semi-analytically, which agrees remarkably well with exact solutions across a wide domain of drive amplitudes spanning many orders of magnitude. Being set up in the framework of Keldysh coherent states path integrals, our approach opens the possibility of studying quantum activation in many-body systems where other approaches are inapplicable.
\end{abstract}

\maketitle

\section{Introduction}

Quantum activation is a process in which a driven system, such as a nonlinear oscillator, switches between two metastable states of forced vibrations due to random noise from spontaneous emission events\cite{Dykman88, marthaler2006switching}. Since energy is continually pumped into the system, these switching events can occur even when the temperature of the bath is zero, unlike the process of classical activation. Although the system may spend the majority of its time close to metastable states, emission events may cause rare fluctuations, taking the system away from one metastable steady-state towards an intermediate unstable state, and then into the basin of attraction of the other metastable state.

The instanton approach in quantum field theory has already been used for evaluating the decay time of metastable states \cite{Kleinert09} and switching rates \cite{Caroli81}, and existing theory can also explain some universal dependencies for switching of a single oscillator close to the bifurcation points in the semi-classical regime \cite{dykman2007critical,Dykman12}. However, significant recent advances in superconducting, electromechanical and optomechanical devices running beyond this regime necessitate a general approach \cite{PhysRevLett.118.040402,brookes_critical_2021}.

\begin{figure*}[!htb]
\includegraphics[width=\linewidth]{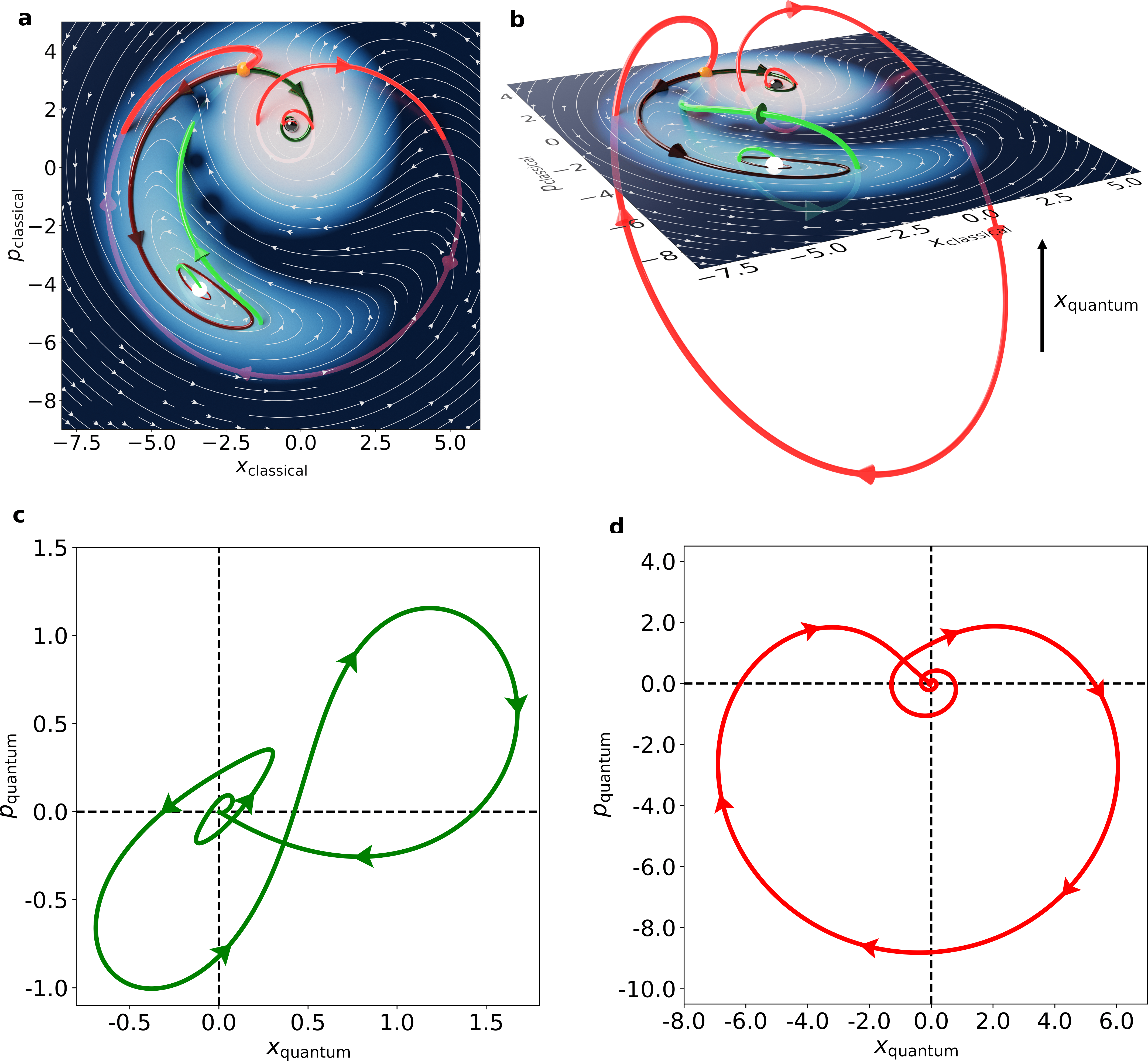}
\caption{\textbf{Keldysh escape paths.} The switching paths following the equations of motion derived using Keldysh field theory. The parameters of the model are $(\chi/\kappa,\delta/\kappa,\epsilon/\kappa) = (-0.5,5.8,-4.0)$. The full four dimensional path cannot be shown, so in panels \textbf{a} (top view) and \textbf{b} (angled view) we plot the two classical variables $x_\mathrm{classical}$ and $p_\mathrm{classical}$ as well as the quantum variable $x_\mathrm{quantum}$. Within the classical plane defined by $x_\mathrm{quantum} = p_\mathrm{quantum} = 0$ we display a density plot of the Wigner function of the steady state along with the fixed points (white, black and yellow) of the classical equations of motion.  The white and black balls mark the bright and dim fixed points respectively, while the yellow ball the unstable point. The switching paths originate at the stable fixed points and immediately leave the classical plane. By utilizing the quantum degrees of freedom, the system is able to escape the classical basin of attraction of the bright state and arrive at the unstable point, from which it may relax classically to the other stable point. The quantum components $x_\mathrm{quantum}$ and $p_\mathrm{quantum}$ of the escape paths from \textbf{c} the bright state and \textbf{d} the dim state start and end at the values of zero, indicating that the escape path starts and ends in the classical plane.} 
\label{fig:3d_switching_path}
\end{figure*}

The method for calculating switching which we now explore is based on Keldysh field theory \cite{elgart_rare_2004,Kamenev11,Torre13,Sieberer16}. Whereas in thermal equilibrium we can obtain switching rates by studying the dynamics of a system in imaginary time \cite{miller1975semiclassical,hanggi1990reaction,wolynes1987imaginary,cao1995computation}, this is no longer possible when drive and dissipation are included. In this situation the state of the system is described by a density matrix and in the formalism of Keldysh field theory this leads to a doubling of the dimensions of the phase space. The additional dimensions open paths for dissipative (noise-based) motion, see Fig. \ref{fig:3d_switching_path}, to be included in the mean-field description of the dynamics and allow us to gain more insight into the processes of activation and switching. With this approach we can semi-analytically calculate the exponential dependence of the switching times on the drive amplitude and frequency, and find that they agree with numerical simulations over a wide range of parameters. In particular, we find our method to work well when the non-linearity of the oscillator is of a similar order as the oscillator decay rate, i.e. outside the semi-classical regime. Below we first describe our analytical approach based on Keldysh field theory, which is general and could be applied in a wide range of scenarios, even including many-particle systems. Later, we will discuss numerical studies of the Lindblad master equation, which we use to validate our analytical results.


\begin{figure*}[!htb]
\includegraphics[width=\linewidth]{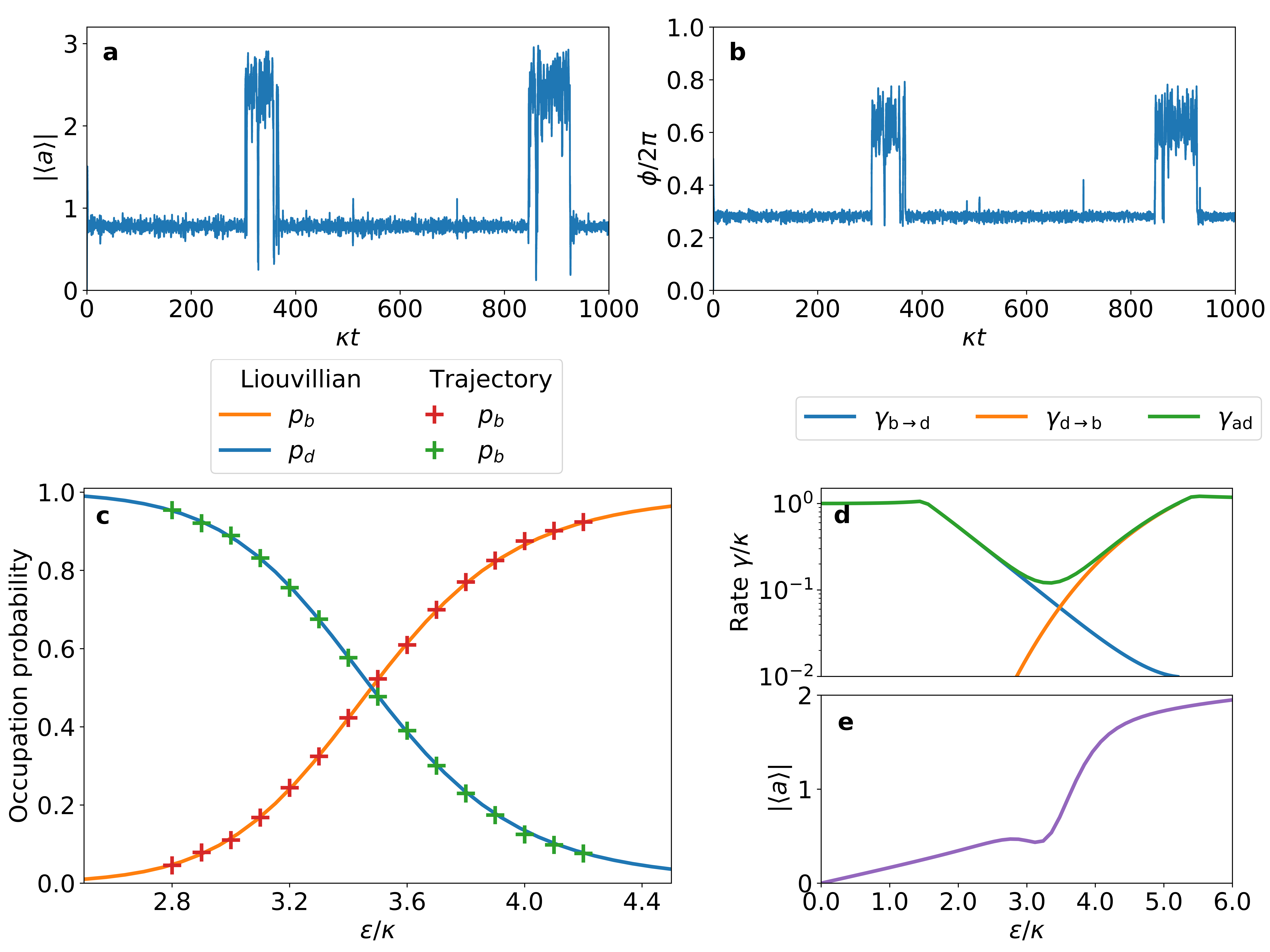}
\caption{\textbf{Results of Liouvillian and quantum trajectory methods.} The amplitude (\textbf{a}) and the phase (\textbf{b}) of the cavity state during a quantum trajectory produced at $(\chi/\kappa,\delta/\kappa,\epsilon/\kappa)=(-0.5,5.8,4.0)$. In the bistable regime this trajectory displays sudden jumps between two metastable states whose lifetimes are typically much longer than the lifetime of the cavity $1/\kappa$. Panel \textbf{c}: the occupation probabilities of the bright and dim states vs the drive amplitude $\epsilon$, obtained by studying the eigenstates of the Liouvillian at $(\chi/\kappa,\delta/\kappa)=(-1.0,6.0)$. As the system moves through the bistable regime it transitions from a state consisting entirely of the dim to entirely of the bright state. The markers indicate occupation probabilities calculated by studying the trajectories produced using a stochastic Schr\"odinger equation. Excellent agreement between these methods is seen. Panel \textbf{d}: the asymptotic decay rate (green)  falls significantly in the bistable regime, indicating the onset of critical slowing down. The bright (blue) and dim (orange) state occupation probabilities  used to calculate the switching rates. Panel \textbf{e}:  the transmission of the cavity as we increase the drive. The bistable regime coincides with a small dip before a sudden increase in the cavity amplitude.} 
\label{fig:master_equation_figure}
\end{figure*}

\begin{figure*}[!htb]
\includegraphics[width=\linewidth]{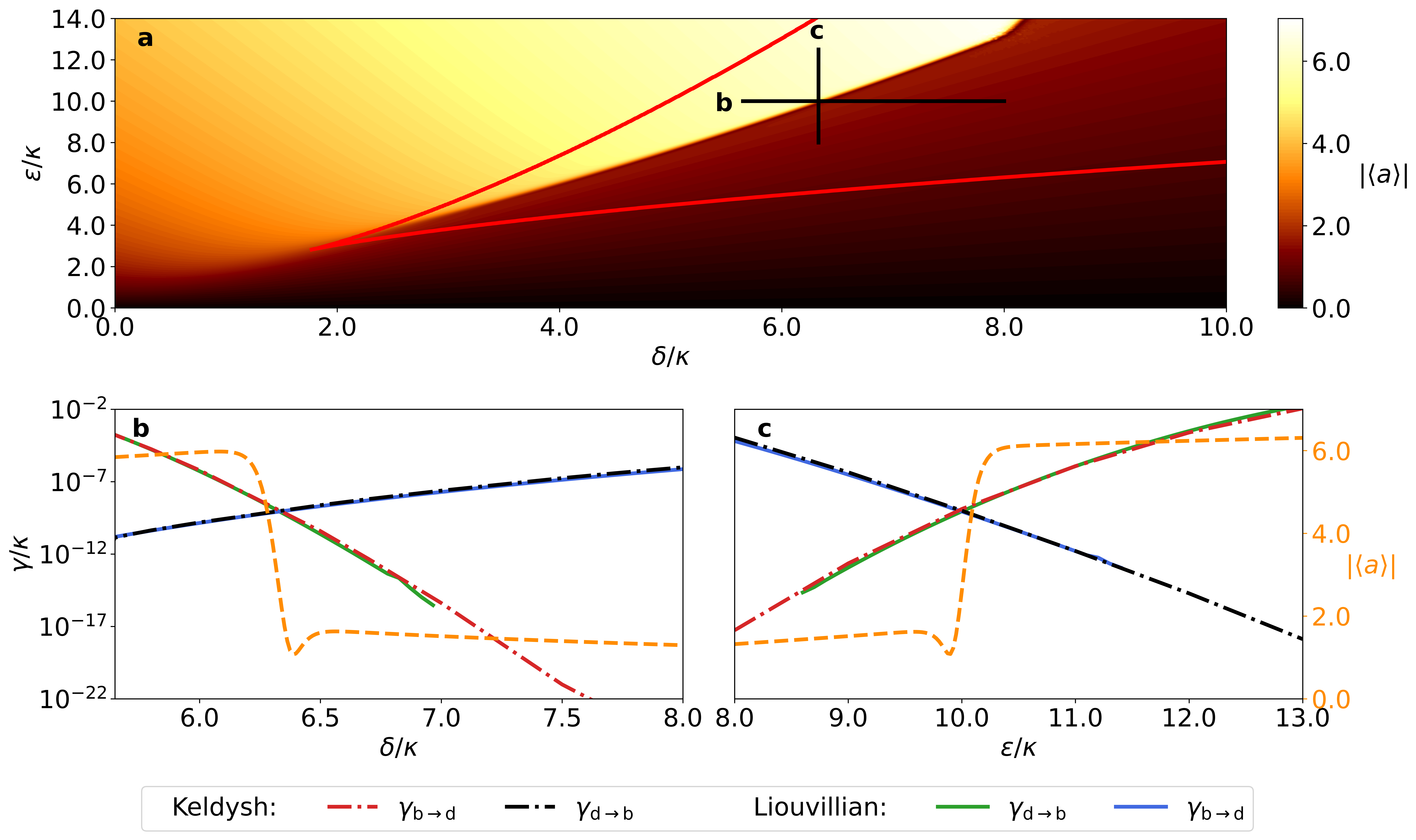}
\caption{\textbf{Comparing Keldysh and Liouvillian switching rates at $\chi/\kappa=-0.1$.} Panel \textbf{a}: the oscillator amplitude as a function of $\delta$ and $\epsilon$. The boundaries of the bistable region (red solid lines) surround a sudden transition from low to high amplitude states as the drive (detuning) is increased (decreased) and are calculated using the classical equations of motion.  In panels \textbf{b} (for $\epsilon/\kappa = 10.0$) and \textbf{c} (for $\delta/\kappa=6.33$) we plot the switching rates and oscillator amplitude along the black lines marked in panel \textbf{a}.  Here, we have fitted $\omega_{\mathrm{b \rightarrow d}}/\kappa = 1.0 $ and $\omega_{\mathrm{d \rightarrow b}}/\kappa = 0.1 $ to give an excellent agreement between the Keldysh and Liouvillian switching rates over several orders of magnitude. This indicates that our assumption that the attempt rates vary slowly with the drive parameters was remarkably accurate and the dominant variations in the switching rates are well described by the action of the optimal switching path. The overlaid cavity amplitude (orange) shows that the crossing of the rates coincides with transition between high and low amplitude states of the oscillator.} 
\label{fig:switching_rates}
\end{figure*}

\section{Dynamics of the Kerr Oscillator}

\subsection{Keldysh Action}

We focus on the prototypical model of the nonlinear driven-dissipative Kerr oscillator in the quantum regime. The self-Kerr effect is a nonlinear shifting of a
resonator frequency as a function of the number of photons in the mode. A simple quantum system where this can been seen is the quantum Duffing oscillator, with its term proportional to $(a^\dagger a)^2$ in the Hamiltonian, where $a$ is the photon annihilation operator for a resonator mode. In the classical limit, this becomes the quadratic dependence of
the refractive index on the electric field strength, sometimes known as self-phase modulation \cite{Drummond_1980, Elliott16, Bishop10}. This effect manifests itself at second order in a series expansion of the Jaynes–Cummings interaction in the dispersive limit \cite{Mavrogordatos17, Elliott_2018,PhysRevB.86.220503,PhysRevA.82.022335}. In a frame rotating at the drive frequency the Kerr oscillator Hamiltonian reads
\begin{equation}
H = \delta a^\dagger a + \chi a^\dagger a^\dagger a a + i \epsilon (a^\dagger - a) \label{eq:quantum_duffing_oscillator_hamiltonian}
\end{equation}
where $\delta$ is the detuning between the oscillator and the drive frequency, $\chi$ is the nonlinearity of the oscillator, $\epsilon$ is the drive amplitude, and we have taken $\hslash=1$. The rotating wave approximation has been applied.

All correlation functions and observables of this system, such as the rate of switching between metastable states, can be obtained from the partition function \cite{Kamenev11,Sieberer16}
\begin{align}
    \mathcal{Z} = \int \mathcal{D}a_{-} \mathcal{D}a^{*}_{-} \mathcal{D}a_{+} \mathcal{D}a^{*}_{+} e^{i S[a_{+},a^{*}_{+},a_{-},a^{*}_{-}]}
\end{align}
which in the Keldysh formalism involves a path integral over two degrees of freedom: the forward $a_+$ and backward $a_-$ time paths. The action S is given by
\begin{align}
S  = &  \!\int\! L~ dt, \\
L = & \,  a_{+}^{*} i \partial_t a_{+} - a_{-}^{*} i \partial_t a_{-} +i\varepsilon \left( a_{+}^{*}-a_{+}-a_{-}^{*}+a_{-}\right)  
 \nonumber \\ & - \delta (a_{+}^{*} a_{+} - a_{-}^{*} a_{-} ) - \chi (a_{+}^{*2} a_{+}^2 - a_{-}^{*2} a_{-}^2 )     \nonumber \\
& - i \kappa \left( 2 a_{+} a_{-}^{*} - a_{+}^{*} a_{+}^{} - a_{-}^{*}a_{-}^{}\right), \label{eq:first_lagrangian}
\end{align}
where $\pm$ denotes the fields in forward/backward branches. The dissipative terms have been obtained from integration over the degrees of freedom of a bosonic bath coupled to the system \cite{Marzena_PRB2007}. In the present case we consider a Markovian bath at zero temperature in order to compare with with numerical results obtained for a Lindblad master equation, but memory effects and finite temperatures could both be included by choosing an appropriate frequency dependent form of $\kappa$.

\subsection{Switching Paths}

We can now define classical and quantum field variables according to $a_c = (a_+ + a_-)/\sqrt{2},~ a_q = (a_+ - a_-)/\sqrt{2}$. In the absence of quantum fluctuations, the evolution will be restricted to the classical plane (i.e. $a_q=0$) for the system to move towards either of two stable fixed points, corresponding to the bright and dim metastable states. Activation and tunnelling events which allow switching from one metastable state to another are expected to occur via the quantum degree of freedom $a_q$.

In order to study the dynamics of our quantum and classical field variables we now decompose them into real and imaginary components according to
\begin{equation}
a_c = (x_c + i p_c)/\sqrt{2}, \quad a_q = (\tilde{x}_q + i \tilde{p}_q)/\sqrt{2}.
\end{equation}
In these terms the Lagrangian reads
\begin{align}
L &= \dot{x_c} \tilde{p}_q - \dot{p_c} \tilde{x}_q - \left[\delta  + \frac{\chi}{2} \left(x_c^2 + p_c^2 + \tilde{x}_q^2 + \tilde{p}_q^2 \right) \right] (x_c \tilde{x}_q + p_c \tilde{p}_q) \nonumber \\
&+  \kappa (x_c \tilde{p}_q - p_c \tilde{x}_q)  + i \kappa (\tilde{x}_q^2 + \tilde{p}_q^2) + 2 \varepsilon \tilde{p}_q
\end{align}
up to total derivatives, while the partition function is now expressed as
\begin{align}
    \mathcal{Z} = \int \mathcal{D}x_c \mathcal{D}p_c \mathcal{D}\tilde{x}_q \mathcal{D}\tilde{p}_q e^{i S[x_c, p_c, \tilde{x}_q, \tilde{p}_q]}. \label{eq:propagator_2}
\end{align}
We may wish to approximate the above Lagrangian by taking into account only the saddle-point (the most probable) paths. This would allow us to determine the equations of motion and find the paths which connect the metastable states, and which dominate the partition function. Unfortunately,  no such paths exist for real values of the co-ordinates (see appendix A). Motion is constrained to the classical plane and there are no paths leaving the metastable states.

At this point we can consider the approach explored in \cite{Kamenev11}: what if the stationary paths lie along the imaginary axes of $\tilde{x}_q$ and $\tilde{p}_q$. instead? Although the integrals in eq. (\ref{eq:propagator_2}) are all along real axes we may use Cauchy's theorem to deform the paths without changing their values. Therefore we choose to make the quantum co-ordinates imaginary and we rewrite them as follows 
\begin{equation}
\tilde{x}_q \rightarrow -i p_q,~ \tilde{p}_q \rightarrow i x_q
\end{equation}
in terms of which the Lagrangian can be written as
\begin{equation}
i L = - \left[\dot{x}_c p_c + \dot{p}_c p_q - H \left(x_c, p_c, x_q, p_q \right)\right]
\end{equation}
with an auxiliary Hamiltonian given by 
\begin{align} \label{eq:keldysh_hamiltonian}
    H &= \Bigg( \delta  + \frac{\chi}{2} \left(x_c^2 + p_c^2 - x_q^2 - p_q^2 \right) \Bigg) (p_c x_q - x_c p_q ) \nonumber \\
& - \kappa (x_c x_q + p_c p_q) + \kappa (x_q^2 + p_q^2) + 2 \varepsilon x_q.
\end{align}
The saddle point equations of motion are then given by
\begin{align} \label{eq:keldysh_equations_of_motion}
\dot{\bm{z}}_c = \partial_{\bm{z}_q} H, \quad\quad \dot{\bm{z}}_q = - \partial_{\bm{z}_c} H
\end{align}
where $\bm{z}_c = (x_c,p_c)$ and $\bm{z}_q = (x_q,p_q)$ and the action, consequently, becomes $iS = i \!\!\int\! L \ dt  =  - \int d \bm{z}_c \cdot \bm{z}_q$. We shall see that our equations of motion now allow evolution out of the classical plane and can connect the metastable states as seen in Fig. \ref{fig:3d_switching_path}.

To identify the fixed points corresponding to these metastable states we obtain the classical equations of motion by setting the quantum variables to zero, $x_q=p_q=0$. It is known that in the bistable regime there are three fixed points within this plane, two of which are stable. We refer to these as the bright and dim states according to the intensity of the oscillator field at those points. Meanwhile, there is a third unstable point lying on the separatrix which divides the plane into the two basins of attraction of the stables states.

Although the bright and dim states are stable within the classical plane, we find that quantum fluctuations can allow rare escape events, during which the system moves to the unstable point along a path lying outside the plane, where quantum components of the fields are non-zero. Once it has reached the unstable point it may relax along the classical path to the other stable point. The trajectories of these escape events are described by the full equations of motion above and examples are displayed in Fig. \ref{fig:3d_switching_path}.

Here we are particularly interested in calculating the rate at which these escape events occur. In the saddle point approximation, the rate of switching from the point $\bm{Z}_j$ to the other stable point $\bm{Z}_k$ can be written as
\begin{equation}
\gamma_{j \rightarrow k} = \omega_{j \rightarrow u} e^{i S_{j \rightarrow u}}
\label{eq:switching_rates}
\end{equation}
where the prefactor $\omega_{j \rightarrow u}$ is the attempt frequency \cite{Kleinert09,Jordan04} and the action is calculated by integrating the Lagrangian along the path from $\bm{Z}_j$ to the unstable point $\bm{Z}_u$. Results from these calculations can be seen in Fig. \ref{fig:switching_rates}\textbf{b} and \ref{fig:switching_rates}\textbf{c}. Details of the calculations can be found in appendix B.

\subsection{Master Equation and Stochastic Schr\"odinger Equation}

Before we discuss these results, we also wish to obtain some exact numerical results via an alternative approach as verification. Assuming the dilute gas limit of instantons \cite{Kleinert09} and an effective two-state model \cite{Jordan04,Roma05}, the occupation probabilities of the bright and dim states are governed by the following rate equation
\begin{equation}\label{eq:me_p}
\frac{d}{dt}
\begin{pmatrix}
{p}_\mathrm{b} \\
{p}_\mathrm{d}
\end{pmatrix}
=
\begin{pmatrix}
-\gamma_{\mathrm{b}\rightarrow \mathrm{d}} & \gamma_{\mathrm{d}\rightarrow \mathrm{b}} \\
\gamma_{\mathrm{b}\rightarrow \mathrm{d}} &  -\gamma_{\mathrm{d}\rightarrow \mathrm{b}}
\end{pmatrix}
\begin{pmatrix}
{p}_\mathrm{b} \\
{p}_\mathrm{d}
\end{pmatrix}.
\end{equation}
At long times the system relaxes to a steady state in which the probabilities are given by
\begin{equation}\label{eq:prob_stationary}
{p}^{ss}_{\mathrm{b}(\mathrm{d})} = \gamma_{\mathrm{d}(\mathrm{b}) \rightarrow \mathrm{b}(\mathrm{d})} / \gamma_\text{total}, \quad \gamma_\text{total} \equiv \gamma_{\mathrm{b}\rightarrow \mathrm{d}} + \gamma_{\mathrm{d}\rightarrow \mathrm{b}}.
\end{equation}
The steady-state occupation probabilities ${p}^{ss}_{\mathrm{b}(\mathrm{d})}$ and the total decay rate $\gamma_\text{total}$ can both be obtained by studying the dynamics of the Liouvillian master equation, as outlined in section C of the appendices. These quantities are plotted in Fig. \ref{fig:master_equation_figure}\textbf{c} and \ref{fig:master_equation_figure}\textbf{d}. Consequently, we are able to obtain the switching rates according to 
\begin{align}
\gamma_{\mathrm{d}(\mathrm{b}) \rightarrow \mathrm{b}(\mathrm{d})} =  {p}_{\mathrm{b}(\mathrm{d})} \gamma_\text{ad}.
\end{align}
In this manner we can calculate the switching rates in the bistable regime, which are also included in Fig. \ref{fig:master_equation_figure}\textbf{d}.

Next, we also obtain switching rates more directly by observing switching events in solutions of the stochastic Schr\"odinger equation for an optical cavity under heterodyne detection \cite{wiseman2009quantum}. By simulating a trajectory over a sufficiently long period of time, we are able to observe many switching events (Fig. \ref{fig:master_equation_figure}\textbf{a} and \textbf{b}) and obtain the occupation probabilities displayed in Fig \ref{fig:master_equation_figure}\textbf{c}, which agree closely with the results of the master equation.

\subsection{Keldysh Switchig Rates}

\begin{figure*}[!htb]
\includegraphics[width=\linewidth]{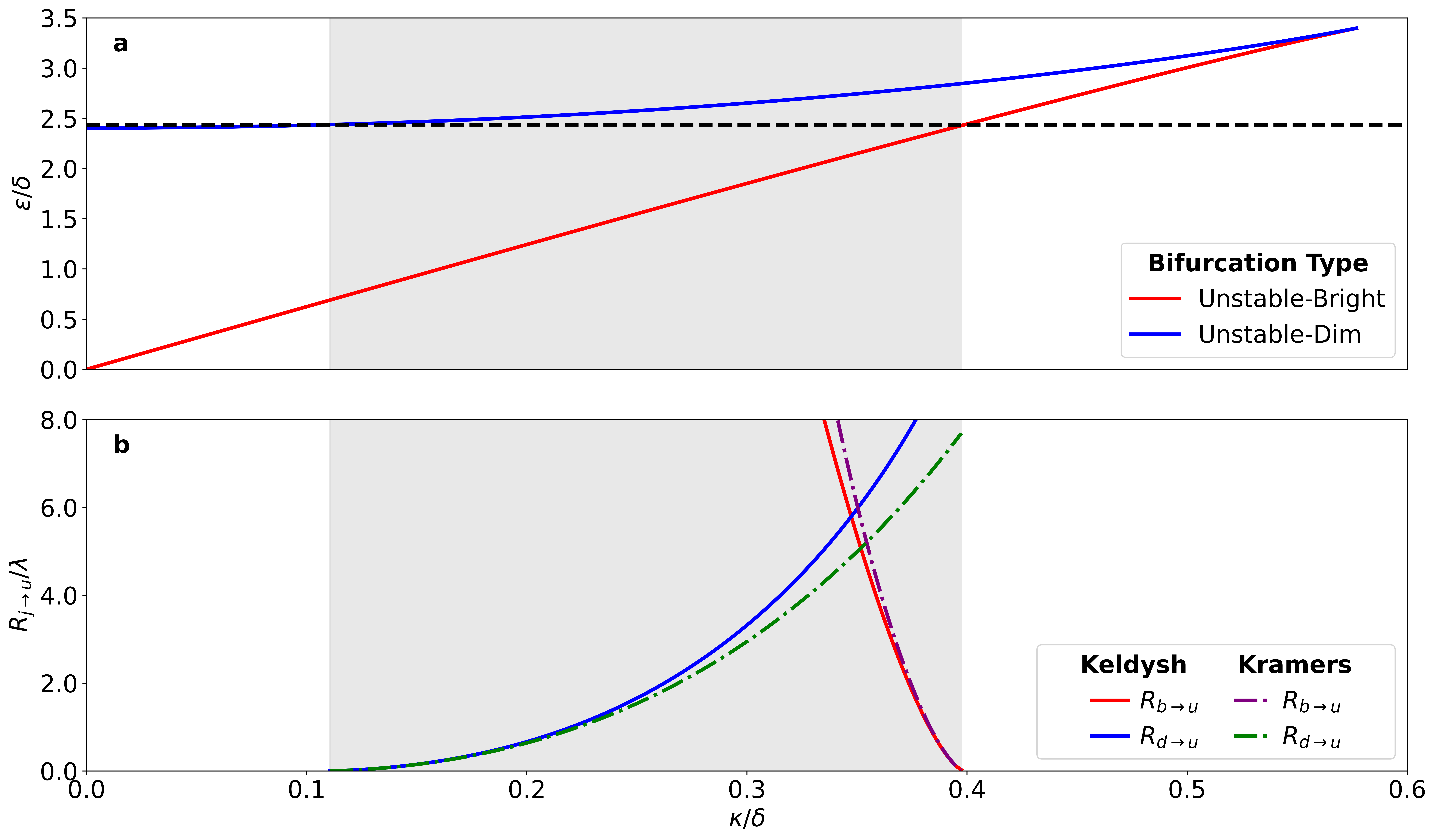}
\caption{\textbf{Measuring the activation barrier height at $\chi/\delta=1/78$.} Panel \textbf{a}: the limits of the bistable regime as a function of $\kappa$ and $\epsilon$. As we increase the drive amplitude the bistable regime appears when the bright and unstable states emerge at the unstable-bright bifurcation. When there is no damping present in the system this bifurcation happens at zero drive. For sufficiently high drive amplitudes the unstable and dim states combine and annihilate each other at the unstable-dim bifurcation. In panel \textbf{b} we plot the variation of the barrier height $R_{j \to u}$ with $\kappa$ at $\epsilon/\delta=2.44$. This barrier height is measured relative to the scaled Planck constant $\lambda = 0.0128$. The barrier height is calculated in two ways. First, we solve the equations of motion derived from the Keldysh approach and calculate the actions along the paths escaping from the metastable states to the unstasble state. These barrier heights are marked by solid lines. Second, we treat escape as a Kramers problem in a one-dimensional potential close to the bifurcation points. These results are marked by dashed lines. Both approaches agree close to the bifurcation points where the one-dimensional approximation can be made, but as begin to disagree in the middle of the bistable regime. Since the switching rates depend exponentially upon the barrier height this can lead to significant disagreements in calculated switching rates, especially at the crossing point where both switching rates and their corresponding populations are equal.} 
\label{fig:barrier}
\end{figure*}

We can now move forward and compare our various methods in Fig. \ref{fig:switching_rates}, using the master equations predictions as a validation for the path integral method. In panel \textbf{a} we first plot the steady-state oscillator amplitude as a function of $\delta$ and $\epsilon$ at $\chi/\kappa=-0.1$. The boundaries of the bistable regime according to the classical equations of motion are marked in red and we see the familiar opening of this regime in the upper right quadrant. Within this regime, we also see the expected sudden transition between low and high amplitude states of the oscillator. These states are separated by a small dip in intensity due to destructive interference between the bistables states, visible in panels \textbf{b} and \textbf{c}.

The black lines in panel \textbf{a} highlight the parameter ranges over which panels \textbf{b} and \textbf{c} were produced. In these ranges we calculated the paths escaping from the metastable states to the unstable state, along with their actions, in order to obtain the switching rates in eq. (\ref{eq:switching_rates}). Since we currently don't have a means to calculate the attempt frequencies, $\omega_{b\rightarrow u}$ and $\omega_{d \rightarrow u}$, we use them as fitting parameters and assume they vary little with changes in $\delta$ or $\kappa$. Despite this we are able to obtain excellent agreement between the switching rates produced by the Keldysh and Liouvillian methods over several orders of magnitude. This indicates that the exponential dependence on the action is by far the dominant factor governing the switching rates and that it can be accurately calculated using the Keldysh method.

Finally, we explore how the switching rates vary with the rate of dissipation $\kappa$. Previous approaches to calculating switching rates have been limited either to the weak dissipation limit \cite{Dykman88,marthaler2006switching,peano2014quantum}, or close to the bifurcation points \cite{dykman2007critical}. In particular, \cite{Dykman88} showcased an approach that involved placing the Lindblad master equation in a co-ordinate representation and applying the WKB approximation to obtain 4-dimensional equations of motion similar to those obtained from the Keldysh approach in eq. (\ref{eq:keldysh_equations_of_motion}). However these equations were only solved in the zero damping limit, which we now extend beyond.

In later work, the switching dynamics were studied in great detail in the vicinity of the bifurcation points \cite{dykman2007critical,Dykman12}. In this regime, the unstable state can be found very close to either of the metastable states and, as these two states approach each other in the phase space, a soft mode emerges between them along which the evolution of the system slows down. The system becomes effectively one-dimensional and the dynamics resemble a Kramers problem in which fluctuations may allow the system may escape from the metastable state by climbing a potential barrier whose peak is found at the unstable state. Beyond this point the system then decays to the other metastable state and the switching event is complete. This approach proved successful and was able to accurately model the switching rates, but only in the vicinity of the bifurcation points where this soft mode emerges.

In Fig. \ref{fig:barrier} we compare the Kramers problem approach with the results of the Keldysh approach. In the Kramers approach the switching rates can be modelled in terms of a barrier height $R_{j \to u}$ as
\begin{equation}
    \gamma_{j \to k} \propto \exp(R_{j \to u} / \lambda)
\end{equation}
In this equation $R_{j \to u}$ represents the barrier height to move from metastable state $j$ to the unstable point $u$. It is rescaled by the scaled Planck constant $\lambda = \hbar \chi / \delta$. In terms of the switching actions we have studied so far, this barrier height is given by $R_{j \to u} = i \lambda S_{j \to u}$. Both the Kramer and Keldysh approaches agree in the vicinity of the bifurcation points where the escape problem becomes one-dimensional. The barrier starts at zero at these bifurcation points and increases as we move into the bistable regime. Both the Kramers and Keldysh approaches show a monotonic dependence on $\kappa$ but disagree quantitatively towards the middle of the bistable regime where the rates are balanced.

\section{Conclusion}

In conclusion, we have shown that the Keldysh technique can be used to obtain an extended mean-field theory, which captures the quantum activation and switching dynamics in a Kerr oscillator. By assuming that the system predominantly follows the saddle point path of the Keldysh action we are able to predict switching rates which are in excellent agreement with numerical simulations of the exact dynamics and we explain how the system moves in the extended classical-quantum phase space between the different fixed points.

The potential of our method goes far beyond this demonstration. In this work we chose to highlight the power of the Keldysh approach when applied to a system in contact with a Markovian thermal bath, which allowed us to crosscheck our results against those obtained using a standard Liouvillian master equation. However, in the future it will be possible to take memory effects into account by, for example, allowing $\kappa$ in eq. (\ref{eq:first_lagrangian}) to vary as a function of frequency. In addition, choosing an appropriate functional form of $\kappa$ will also allow us to study the effects of finite bath temperatures on the process of quantum activation.

Furthermore, since the Keldysh approach is formulated in the language of second quantisation (i.e. coherent state path integrals) it can be straightforwardly applied to more complex many-body systems such as coupled oscillators, spins coupled to bosons or even bosonic lattices. Although methods already exist, which can produce similar results for the Kerr oscillator, it would not be possible to apply them to more complex systems. For example, an alternative approach to the calculation of switching rates has previously been explored based on applying the WKB approximation to the evolution of the Wigner function \cite{Dykman88}. As in the Keldysh method, this approach relies on calculating the action along a path escaping from the basin of attraction via the unstable point. In the limit of weak nonlinearity the equations of motion in the WKB and Keldysh methods converge, however the WKB method would only be applicable to a single particle moving in an external potential and it would be challenging to extend to non-Markovian dynamics. This generalisability is a key advantage of the Keldysh approach and will be the main theme of future work as we go beyond regimes which can easily be compared with other methods.

C.L. acknowledges support by National Research Foundation of Korea (Grant no. NRF-2017R1D1A1B04032142). M.H.S. acknowledges support from EPSRC grant EP/S019669/1, EP/R04399X/1 (Quantera InterPol), EP/K003623/2. E.G. acknowledges support from EPSRC grant EP/I026231/1.

\section{Appendices}

\subsection{Equations of motion}
In terms of field quadratures the Lagrangian is given by
\begin{align}
L &= \dot{x_c} \tilde{p}_q - \dot{p_c} \tilde{x}_q +  \kappa (x_c \tilde{p}_q - p_c \tilde{x}_q) \nonumber \\
&- \left[\delta  + \frac{\chi}{2} \left(x_c^2 + p_c^2 + \tilde{x}_q^2 + \tilde{p}_q^2 \right) \right] (x_c \tilde{x}_q + p_c \tilde{p}_q) \nonumber \\
&  + i \kappa (\tilde{x}_q^2 + \tilde{p}_q^2) + 2 \varepsilon \tilde{p}_q 
\end{align}
while the partition function is given by
\begin{align}
    \mathcal{Z} = \int \mathcal{D}x_c \mathcal{D}p_c \mathcal{D}\tilde{x}_q \mathcal{D}\tilde{p}_q e^{i S[x_c, p_c, \tilde{x}_q, \tilde{p}_q]}
\end{align}
We may wish to estimate this partition function using a saddle point approximation. In this method we find the dominant path in the integral above by solving the Euler-Lagrange equations for the Lagrangian above. These equations of motion are given by
\begin{align}
    \partial_t x_c =& \delta p_c - 2 \epsilon - 2 i \kappa \tilde{p}_q - \kappa x_c \nonumber \\
    &+ \frac{1}{2} \chi (p_c^3 + 3 p_c \tilde{p}_q^2 + p_c x_c^2 + p_c \tilde{x}_q^2 +2 \tilde{p}_q x_c \tilde{x}_q) \\
    \partial_t p_c =& - \delta x_c - \kappa (p_c - 2 i \tilde{x}_q) \nonumber \\
    &- \frac{1}{2} \chi (p_c^2 x_c + 2 p_c \tilde{p}_q \tilde{x}_q + \tilde{p}_q^2 x_c + x_c^3 + 3 x_c \tilde{x}_q^2) \\
    \partial_t \tilde{x}_q =& \frac{1}{2} \chi (3 p_c^2 \tilde{p}_q + 2 p_c x_c \tilde{x}_q + \tilde{p}_q^3 + \tilde{p}_q x_c^2 + \tilde{p}_q \tilde{x}_q^2 ) \nonumber \\ &+ \delta \tilde{p}_q + \kappa \tilde{x}_q \\
    \partial_t \tilde{p}_q =& -\frac{1}{2} \chi (p_c^2 \tilde{x}_q + 2 p_c \tilde{p}_q x_c + \tilde{p}_q^2 \tilde{x}_q + 3 x_c^2 \tilde{x}_q + \tilde{x}_q^3) \nonumber \\ &- \delta \tilde{x}_q + \kappa \tilde{p}_q
\end{align}
If we examine these equations carefully it becomes clear that there are no solutions for purely real values of $x_c$, $p_c$, $\tilde{p}_q$ and $\tilde{p}_q$. If all co-ordinates are initialised to real values then $x_c$ and $p_c$ will immediately evolve to complex values. However we may be able to find solutions where $x_c$ and $p_c$ are both real while $\tilde{p}_q$ and $\tilde{p}_q$ are purely imaginary.

This may seem problematic since the path integrals in the partition function above are over real values of $\tilde{p}_q$ and $\tilde{p}_q$. However, we can use Cauchy's theorem to deform the contours of integration to follow the imaginary axes without changing the values of the integrals \cite{Kamenev11}.
To account for this we find it convenient to rewrite the equations of motion using
\begin{equation}
\tilde{x}_q \rightarrow -i p_q,~ \tilde{p}_q \rightarrow i x_q
\end{equation}
In terms of these real co-ordinates we find
\begin{align}
    \partial_t x_c =& \delta p_c - 2 \epsilon + 2 \kappa x_q - \kappa x_c \nonumber \\
    &+ \frac{1}{2} \chi (p_c^3 - 3 p_c x_q^2 + p_c x_c^2 - p_c p_q^2 +2 x_q x_c p_q) \\
    \partial_t p_c =& - \delta x_c - \kappa (p_c - 2 p_q) \nonumber \\
    &- \frac{1}{2} \chi (p_c^2 x_c + 2 p_c x_q p_q - x_q^2 x_c + x_c^3 - 3 x_c p_q^2) \\
    \partial_t x_q =& \frac{1}{2} \chi (p_q p_c^2 - p_q^3 + 3 p_q x_c^2 - p_q x_q^2 - x_c p_c x_q) \nonumber \\ &+ \delta p_q + \kappa x_q \\
    \partial_t p_q =& \frac{1}{2} \chi (2 p_q p_c x_c + x_q p_q^2 + x_q^3 - x_q x_c^2 - 3 p_c^2 x_q) \nonumber \\ &- \delta X_q + \kappa p_q
\end{align}
The solution to the above equations of motion can be used in the saddle-point approximation of the partition function. For cases where $\kappa$ becomes frequency dependent, e.g. at finite temperature, the action will be non-local in time and the saddle-point equations will become a set of coupled integro-differential equations.

\subsection{Obtaining switching paths}

The equations of motion (\ref{eq:keldysh_equations_of_motion}) have three fixed points within the classical plane, two of which are stable while the other is unstable. The dim and bright fixed points are denoted by $\bm{Z}_d$ and $\bm{Z}_b$ respectively and the unstable point is denoted by $\bm{Z}_u$. The two stable points each have their own basin of attraction and the unstable point lies on the boundary which separates these two basins. In order to switch from one stable point to another the system must leave the classical plane by utilizing the quantum dimensions $x_q$ and $p_q$. The path of least action takes the system to the unstable point, from which it may move into the basin of attraction of the other stable point.

In order to classify the fixed points we linearise the equations of motion around them. At the stable points we find two eigenvalues given by $-\kappa \pm i \omega$ with eigenvectors residing in the classical plane. This indicates that these points are stable within the plane. On the other hand, these points also have two eigenvectors with nonzero quantum components and eigenvalues $\kappa \pm i \omega$, so they are not stable when we consider the full four-dimensional space.


As for $\bm{Z}_u$, the eigenvectors corresponding to eigenvalues $-\kappa_1$ and $\kappa_2$ ($\kappa_1 > \kappa_2 >0$) reside in the classical plane, which indicates that it is saddle point within the classical plane. Meanwhile, the eigenvectors corresponding to eigenvalues $\kappa_1$ and $-\kappa_2$ have nonzero quantum components. The fact that the eigenvalues of fluctuation eigenvectors are nothing but negative of deterministic ones is characteristic of fluctuation-induced escape mechanism \cite{Smelyanskiy97}.



The probability of a successful escape event is proportional to $e^{i S_{j \rightarrow u}}$ where $i S_{j \rightarrow u}$ is the action calculated along the path from stable fixed point $j$ to the unstable point, and is given by
\begin{equation}
iS_{j \rightarrow u} = - \!\int_{\bm{Z}_j \rightarrow \bm{Z}_u}\!\!\!\!\!\!  d{\bm z}_c \cdot \bm{z}_q,
\end{equation}
This action is computed from the numerical solution of the set of equations of motion. By the previous linearisation analysis around $\bm{Z}_u$, we know there is a negative eigenvalue $-\kappa_2$ and the corresponding eigenvector having nonzero quantum values. We use this eigenvector $\bm v$ and obtain the bounce solutions \cite{Coleman85} of the equations of motion with two initial points $\bm{Z}_u \pm \Delta \bm{v}$. We set $\Delta$ as small as our computing system allows. This bounce solution consists of initialising the system at the saddle point and integrating the equations of motion backwards in time until the system reaches one of the stable points. We have numerically confirmed that two solutions evolve to the two stable points respectively and calculated the escape actions according to the integral above.

These solutions have been crosschecked by treating the task of obtaining switching paths as a boundary value problem. In this treatment we apply two boundary conditions at each end of the path. Close to the stable fixed point we constrain the system to occupy the plane spanned by the outgoing eigenvectors whereas at the unstable point we constrain the system to occupy the plane spanned by the incoming eigenvectors. The path connecting these initial and final conditions is then obtained using the scipy.integrate.solve\_bvp function in SciPy \cite{2020SciPy-NMeth}.

\subsection{Extracting bright and dim states from Liouvillian eigenvectors}

In the following we will show how the relaxation rate $\gamma_\text{total}$ and stationary probabilities ${p}^{ss}_{\mathrm{b}(\mathrm{d})}$ can be calculated by studying the Lindblad master equation directly. This provides an alternative method for calculating the switching rates which can then be compared with the results of the Keldysh formalism. The master equation in question is given by

\begin{equation}
    \partial_t \rho = - i [H, \rho] + \kappa \big(a \rho a^\dagger - a^\dagger a \rho - \rho a^\dagger a \big)
\end{equation}
Since the master equation is linear, we can rewrite the evolution in terms of the Liouvillian superoperator $\mathcal{L}$ \cite{minganti2018spectral}
\begin{equation}
    \partial_t \rho = \mathcal{L} \rho.
\end{equation}
The eigenvalue equation of this operator takes the form $\mathcal{L} \rho_m = -(\gamma_m + i\omega_m) \rho_m$, where the real and imaginary components of the complex eigenvalues are denoted by $\gamma_m$ and $\omega_m$ respectively. We can write down the evolution of a state in this eigenbasis as
\begin{align}
\rho(t) = \sum_{m} c_m e^{- (\gamma_m + i \omega_m) t } \rho_m.
\end{align}
We see that $\gamma_m$ represents the decay rate of the component $\rho_m$ and $\omega_m$ represents its oscillation frequency. It is known that $\gamma_m \geq 0$ for all eigenvectors \cite{minganti2018spectral} and this ensures that $\rho$ is well-behaved at long times. States for which $\gamma_m > 0$ will decay over time until the only remaining components of $\rho(t)$ consists of those eigenvectors for which $\gamma_m=0$. For our system we expect a single such eigenvector which forms the steady-state, denoted by $\rho_{\mathrm{ss}}$. In the bistable regime this state will consist of a mixture of the two metastable states, as we can see in the Wigner function displayed in Figs. \ref{fig:3d_switching_path}\textbf{a} and \ref{fig:3d_switching_path}\textbf{b} and in the trajectory displayed in Fig. \ref{fig:master_equation_figure}\textbf{a} and \textbf{b}:
\begin{align} \label{eq:steady_state_structure}
    \rho_{\mathrm{ss}} = p_\mathrm{b} \rho_\mathrm{b} + p_\mathrm{d} \rho_\mathrm{d}.
\end{align}
However we are also interested in the asymptotically decaying eigenvector, i.e. the eigenvector with the smallest finite value of $\gamma_m$, which will be denoted by $\rho_{\mathrm{ad}}$. At long times the state of the system will consists of a mixture of the steady-state and this asymptotically decaying eigenvector, all other eigenvectors having already decayed to negligible levels.

We now have two alternative descriptions of the transient response of the system: one from the Keldysh approach and one from the Liouvillian approach. The Keldysh approach shows us that the system approaches steady-state via switching events between the two metastable states which eventually cause the system to reach a dynamic equilibrium whereby the rates in each direction are balanced. This equilibration occurs at the rate described in eq. (\ref{eq:prob_stationary}). But now we see that this process is also described by the decay of an unknown asymptotically decaying eigenvector at rate $\gamma_{\mathrm{ad}}$. These rates are in fact identical, i.e. $\gamma_{\mathrm{ad}} = \gamma_{\mathrm{total}}$, and the asymptotically decaying state represents imbalance of the occupation probabilities of these states from the eventual steady-state. It can be written as
\begin{equation} \label{eq:ad_state_structure}
    \rho_{\mathrm{ad}} = N(\rho_\mathrm{d} - \rho_\mathrm{b})
\end{equation}
where the normalization $N$ can be set by taking $\Tr(\rho_{\mathrm{ad}}^2) =1$.

Given that the steady and asymptotically decaying eigenvectors are linearly independent mixtures of bright and dim states, we might consider how we can combine them to isolate their components and the corresponding occupation probabilities. These could then be used to calculate the switching rates in eq. (\ref{eq:prob_stationary}).

\begin{figure}
\includegraphics[width=\columnwidth]{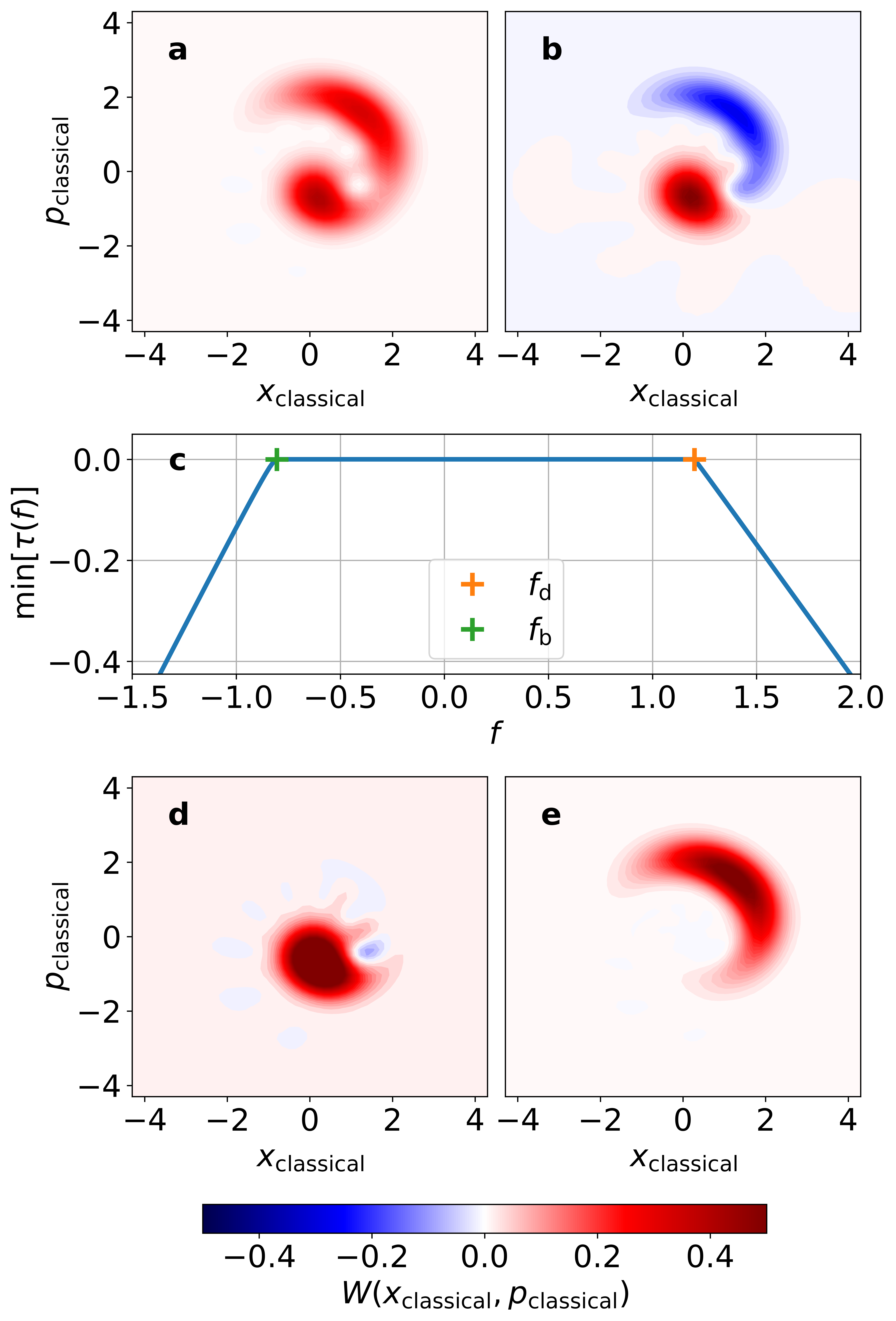}
\caption{\textbf{Extracting the metastable states from the Liouvillian.}  Wigner functions of the steady state (panels \textbf{a}) and the asymptotically decaying eigenvector (panels \textbf{b}) produced at $(\chi/\kappa,\delta/\kappa,\epsilon/\kappa) = (1.0,6.0,3.6)$. Steady state is a mixture of the two metastable states while the asymptotically decaying eigenvector is an antisymmetric mixture. 
This allows to reconstruct the metastable states from a sum of steady and asymptotically decaying eigenvectors. Panel \textbf{c}: the smallest eigenvalue of $\tau(f) = \rho_{\mathrm{ss}} + f \rho_{\mathrm{ad}}$. At $f=f_\mathrm{b}$ and $f=f_\mathrm{ d }$ the state $\tau(f)$ consists entirely of the dim (panels \textbf{d}) and bright (\textbf{e}) states respectively. 
\label{fig:demonstrate_liouvilian_method}}
\end{figure}

In Fig. \ref{fig:demonstrate_liouvilian_method} we illustrate the method of extracting the bright and dim states from the eigenstates of the Liouvillian superoperator describing the evolution of the state at $(\chi/\kappa,\delta/\kappa,\epsilon/\kappa) = (1.0,6.0,3.6)$. We first display the Wigner functions of \textbf{a} the steady state $\rho_{ss}$ and \textbf{b} the asymptotically decaying eigenvector $\rho_{\mathrm{ad}}$ and we observe that these components both consist of a weighted sum of bright and dim states according to the structure outlined in eqs. (\ref{eq:steady_state_structure}) and (\ref{eq:ad_state_structure}). In the steady state the weights correspond to probabilities and are both positive, resulting in the two peaks observed in panel \textbf{a}. Meanwhile in the asymptotically decaying eigenvector the weights are equal in magnitude but opposite in sign, which results in the peak and dip seen in panel \textbf{b}.

These two states can be combined to form the mixture $\tau$ defined by
\begin{align}
\label{eq:definition_of_tau}
\tau(f) &= \rho_{\mathrm{ss}} + f \rho_{\mathrm{ad}} \nonumber \\
&= (p_\mathrm{b} - f N) \rho_\mathrm{b} + (p_\mathrm{d} + f N) \rho_\mathrm{d}.
\end{align}
In order to extract the bright and dim states we plot the minimum eigenvalue $\mathrm{min}(\tau)$ against $f$ in panel \textbf{c} and identify the points at which this eigenvalue falls below zero. Finally we plot the Wigner functions of the resulting bright and dim states. In panel \textbf{d} we display $\rho_\mathrm{d} \propto \tau(f_\mathrm{d})$ while in panel \textbf{e} we display $\rho_\mathrm{b} \propto \tau(f_\mathrm{b})$. We see that these states consist of a single peak, as expected.

In order that $\rho_\mathrm{b}$ and $\rho_\mathrm{d}$ are both physically realistic states they should be positive semidefinite, i.e. they should have no negative eigenvalues. If we define the function $\mathrm{min}(\tau)$, which returns the smallest eigenvalue of $\tau$, then our condition can now be stated as $\mathrm{min}(\rho_\mathrm{b}),\mathrm{min}(\rho_\mathrm{d}) \geq 0$.
Next we assume the metastable states do not overlap, i.e. $\Tr ( \rho_\mathrm{b} \rho_\mathrm{d} ) = 0$, which is a good approximation provided the drive amplitude is sufficiently strong for the bistable states to be well separated. Given this assumption, the state $\tau(f)$ will be positive semidefinite if and only if the coefficients of the bistable states are both greater than or equal to zero. Therefore, we can identify the values $f_\mathrm{ d }=p_\mathrm{b} / N$ and $f_\mathrm{b}=-p_\mathrm{d} / N$ by plotting $\mathrm{min}[\tau(f)]$ as a function of $f$ and locating where this function falls below zero. The values of $f_\mathrm{b}$ and $f_\mathrm{ d }$ thus obtained are then combined with the normalization $p_b + p_d = 1$ to obtain the occupation probabilities
\begin{align}
p_{\mathrm{b}} = \frac{f_\mathrm{b}}{f_\mathrm{b}-f_\mathrm{ d }}, \quad\quad p_{\mathrm{d}} = -\frac{f_\mathrm{ d }}{f_\mathrm{b}-f_\mathrm{ d }}
\end{align}
which are plotted in Fig. \ref{fig:master_equation_figure}\textbf{c}.

\subsection{Comparison with WKB}

The switching between metastable states of the Duffing oscillator has previously been studied using other techniques, such as the WKB method. These studies start from a Hamiltonian given by
\begin{equation}
    H_\mathrm{Duffing}(t) = \frac{1}{2} p^2 + \frac{1}{2} \omega_0^2 q^2 + \frac{1}{4} \gamma q^4 - q A \cos(\omega_F t)
\end{equation}
before moving to a rotating frame by transforming to the following variables:
\begin{align}
    q &= C_{res}(Q \cos(\omega_F t) + P \sin(\omega_F t)) \\
    p &= - C_{res} \omega_F (Q \sin(\omega_F t) - P \cos(\omega_F t))
\end{align}
where $C_{res} = (8 \omega_F \delta\omega / 3 \gamma)^{1/2}$ and $\delta\omega = \omega_F - \omega_0$ and the new position and momentum variables follow the commutation relation
\begin{equation}
    [P,Q] = - i \lambda, \quad \lambda = 3 \hbar \gamma / 8 \omega_F^2 \delta\omega.
\end{equation}
We can transform this into the same form as eq. (\ref{eq:quantum_duffing_oscillator_hamiltonian}) using ladder operators according to
\begin{equation}
    Q = \sqrt{\frac{\lambda}{2}}(a + a^\dagger), \quad P = i \sqrt{\frac{\lambda}{2}}(a^\dagger - a). \\
\end{equation}

\bibliography{Keldysh,PathIntegral,Bistability}

\end{document}